\documentclass[aps, prb, twocolumn, 11pt, reprint, floatfix, a4paper, showpacs]{revtex4-1} 
\usepackage{graphicx}
\usepackage{textcomp} %required to provide \textmu, a lower case non-italic mu used in for "micro" 
\PassOptionsToPackage{caption=false}{subfig}
\usepackage{subfig}

\begin{document}

\title{Quantum Hall induced currents and the magnetoresistance of a quantum
point contact}

\author{M. J. Smith}\author{C. D. H. Williams}\author{A. Usher}

\affiliation{School of Physics, University of Exeter, Stocker Road, Exeter EX4 4QL}

\author{A. S. Sachrajda}\author{A. Kam}\author{Z. R. Wasilewski}

\affiliation{Institute for Microstructural Sciences, National Research Council of Canada, Ottawa, Ontario K1A 0R6, Canada}

\date{\today}

\begin{abstract}
We report an investigation of quantum Hall induced currents by simultaneous measurements of their
magnetic moment and their effect on the conductance of a quantum
point contact (QPC). Features in the magnetic moment and QPC
resistance are correlated at Landau-level filling factors $\nu=1,2$
and $4$, which demonstrates the common origin of the effects. Temperature and non-linear sweep rate dependences are observed to be similar for
the two effects.  Furthermore, features in the noise of the induced currents, caused by breakdown of the quantum Hall effect, are observed to have clear correlations between the two
measurements. In contrast, there is a distinct difference in the way
that the induced currents decay with time when the sweeping field
halts at integer filling factor.  We attribute this difference to the fact that, while both effects are sensitive to the magnitude of the induced current, the QPC resistance is also sensitive to the proximity of the current to the QPC split-gate.  Although it is clearly demonstrated that induced
currents affect the electrostatics of a QPC, the reverse effect, the QPC influencing the induced current, was not observed.
\end{abstract}


\maketitle
\section{Introduction}
The occurrence of long-lived induced currents when a two dimensional electron system
(2DES) is in the quantum Hall effect (QHE) regime demonstrates the extraordinarily low dissipation accompanying the effect (for a review see [\onlinecite{Usher2009}]).
Induced currents are expected to be present in conventional quantum
Hall measurements, but are not detectable in such experiments despite
often being orders of magnitude larger than the currents injected
into the 2DES through the electrical contacts. This isolation of the induced currents from the edge states that cause the conventional QHE has led to the suggestion\cite{Klaffs2004} that the induced currents flow within the innermost of the incompressible strips proposed by \citet{Chklovskii1992} However, we recently discovered
that induced currents \textit{can} influence conventional transport measurements in electrostatically
defined nanostructures such as quantum point contacts (QPCs) or quantum
dots.\cite{Pioro-Ladriere2006} Features were observed in the conductance of a QPC and the Coulomb
blockade spectrum of a quantum dot which were hysteretic with respect to magnetic-field sweep direction. In the same work, separate experiments measuring the magnetic
moment associated with the QHE induced currents showed that they were present in the same GaAs heterojunction from which the nanostructures were fabricated. The filling-factor, temperature, and
sweep-rate dependences of the induced currents in the heterojunction correlated with the hysteretic
behaviour observed within the nanostructures, suggesting that a common origin was possible.
In this paper we report the first \emph{simultaneous} measurements of induced currents
from their magnetic moments and their effects on QPC conductance, confirming their common origin and demonstrating that the QPC resistance is sensitive not only to the magnitude of the induced current but also to its distance from the QPC split-gate.
\begin{figure}
\includegraphics[clip=true, width=0.41\textwidth]{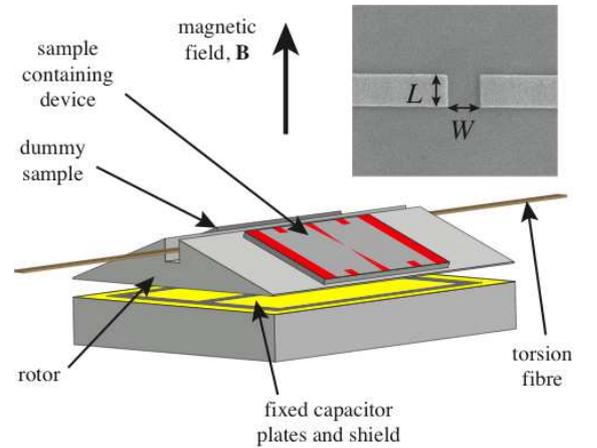}
\caption{Schematic diagram of the QPC device mounted on the torsion-balance magnetometer. The magnetic moment $\mathbf{m}$ caused by the induced currents circulating in the 2DES `leads' of the device produces a torque, $\mathbf{m}\times\mathbf{B}$.  This is detected as an imbalance of the differential capacitor formed by an electrode on the underside of the rotor and the two fixed capacitor plates shown.  Simultaneously, the QPC conductance is measured; electrical connections to the device are not shown on the diagram, for clarity. The inset is a scanning electron microscope image of the QPC split-gate, $W=509\,$nm and $L=503\,$nm.\label{fig:sample}}
\end{figure}
\section{Experiment details}
The device was fabricated at the Institute for Microstructural
Sciences,\cite{AK47} from a GaAs--(Al,Ga)As heterojunction, forming
a 2DES of dimensions $4.0\,\mathrm{mm}$ by $4.5\,\mathrm{mm}$ with
a metallic split-gate defined on the surface, approximately 110$\,$nm
above the 2DES. The split gate was centred over the 2DES.
The lithographic dimensions of the gate are given in Fig.~\ref{fig:sample}.
The QPC was defined electrostatically by the application of a
negative bias, with respect to the 2DES, to both sides of the split gate.
Gold ohmic contact pads on the QPC were connected to insulated copper
twisted pairs (25$\,$\textmu m diameter with approximately
3 twists per millimetre) with colloidal-silver paint.\cite{Silverpaint} The device was mounted
on the rotor of a torsion-balance magnetometer (Fig.~\ref{fig:sample}); details are
given elsewhere.\cite{Matthews2004a}  The normal to the 2DES was tilted at an angle of $20^{\circ}$ to the applied magnetic field. The wires from the device were arranged to minimise mechanical perturbation of the magnetometer.
\begin{figure}
\includegraphics[trim=0cm 0cm 0cm 0.55cm, clip=true, width=0.47\textwidth]{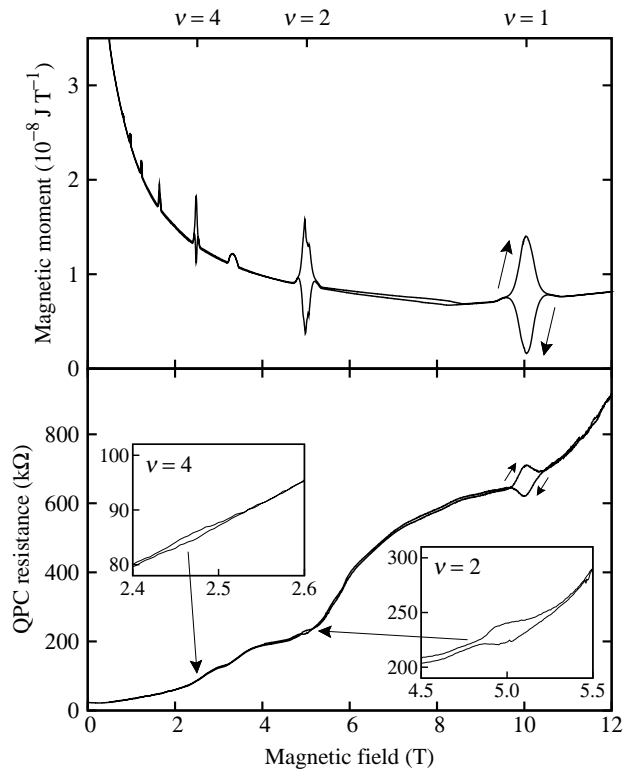}

\caption{Simultaneous measurement of magnetic moment (upper trace) and QPC resistance (lower trace) at $300\mathrm{\, mK}$.  The magnetic-field sweep rate was $1.6\,\mathrm{mT\, s^{-1}}$.  Arrows indicate the magnetic-field sweep directions.  The two insets show the detail of the hysteretic resistance at $\nu=2$ and $\nu=4$.\label{fig:dual_meas}}

\end{figure}

In order to reduce noise, predominantly manifest as telegraph noise,\cite{Pioro-Ladriere2005} the sample was cooled from room temperature with a
gate voltage of $+0.26\,\mathrm{V}$ applied. This removed free charge from the gate region whilst it was still mobile. A standard four-terminal measurement with a low-frequency (17.3~Hz)
lock-in detector was used to determine the QPC resistance, with
an excitation current of $10\,\mathrm{nA}$.
Once cold, the gate voltage was set to a negative value such that conduction through the QPC was
below $2e^{2}/h$, and hence the mechanism for conduction was quantum
tunnelling. In this state the QPC was very sensitive to changes in
its local electrostatic environment. With this arrangement, \textit{simultaneous}
measurements were made of the magnetic moment of the induced current in the regions
of 2DES either side of the QPC, and of the conduction through the QPC.

The magnetometer occupied the mixing chamber of a dilution refrigerator and
measurements were taken between $39\,\mathrm{mK}$ and $1.6\,\mathrm{K}$.
The magnetic field was produced by a $19\,\mathrm{T}$ solenoid, driven
by an Oxford Instruments IPS 120-20 digital power supply.
\section{Results}
\subsection{Simultaneous measurements of magnetic moment and QPC resistance}\label{sub:simultaneous}
Figure~\ref{fig:dual_meas} shows the results of the simultaneous
measurement of the magnetic moment and the QPC resistance. Evidence
of induced currents (features that reverse with magnetic-field sweep
direction) are seen at Landau-level filling factors $\nu=1,2$ and
$4$ in both the magnetic moment and the QPC resistance data. The smaller
features in the magnetic moment at $\nu=3,6,8$ and $10$ are primarily caused by capacitive
coupling between the 2DES and the capacitor plates (the shield referred to in Fig.~\ref{fig:sample} was designed to minimise this coupling), but on close inspection also show a small hysteretic effect. 

The relative sizes of the features $\nu=1,2$ and $4$ for the magnetic
moment are approximately $1.7:1.7:1$, while for the QPC magnetoresistance features they are
$46:4.3:1$.  We suggest two possible explanations for this difference. First, the QPC measurement is influenced not only by the size of the induced current but also, very strongly, by the distance of the current path from the edge.  This distance increases as $B$ decreases (i.e. as the magnetic length $\left({\hbar}/{eB}\right)^{1/2}$ increases), reducing the sensitivity of  the QPC to the induced current.  Second, the sensitivity of the QPC depends
on the background resistance, which is changing with magnetic field
due to a background magnetoresistance.  These two explanations are not mutually exclusive.
\begin{figure}
\includegraphics[trim=0cm 0cm 0cm 0.85cm, clip=true, width=0.47\textwidth]{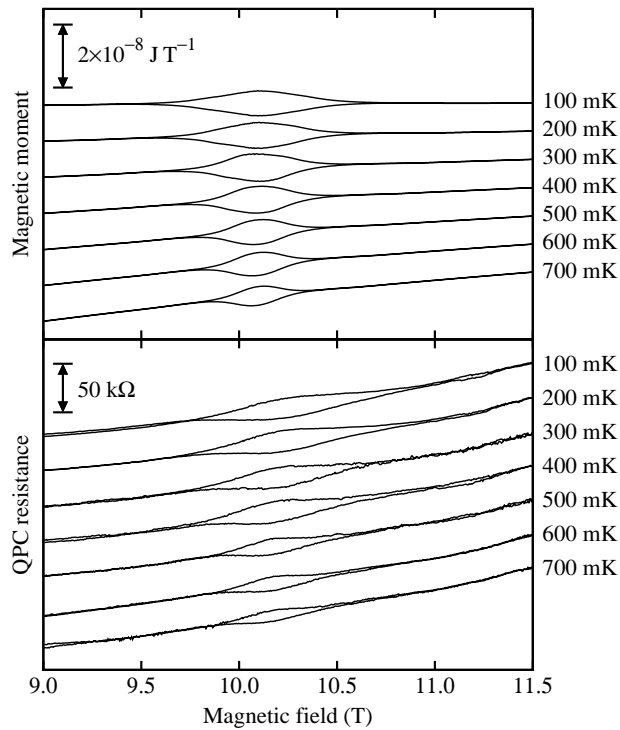}

\caption{Temperature dependence of the induced current at $\nu=1$ measured via magnetic moment (top), and hysteretic QPC resistance (bottom).  Both up- and down-sweeps are shown for each temperature, and each temperature is offset for clarity. The
magnetic field was swept at a rate of $3.2\,\mathrm{mT\, s^{-1}}$. Features in both measurements
are suppressed as temperature is increased.\label{fig:T_depend}}

\end{figure}
\subsection{Temperature dependence}\label{sub:temp}
Figure~\ref{fig:T_depend} shows the temperature dependence of the
two measurements at $\nu=1$ at a relatively high sweep rate of $3.2\,\mathrm{mT\, s^{-1}}$.
The widths of the hysteretic features in both measurements reduce monotonically with increasing temperature, while the heights remain approximately constant
up to a temperature of $400\,\mathrm{mK}$ and then gradually reduce as
the temperature is increased further.  For $\nu=2$ (not shown) the region of constant
feature-height persists up to 800$\,$mK.
There is a significant difference, of about $0.15\,\mathrm{T}$, in
the magnetic fields at which the magnetic moment and resistance features
occur, and there is a significant asymmetry to the resistance curves.
These are not experimental artefacts, and will be discussed below. 
\begin{figure}
\includegraphics[trim=0cm 0cm 0cm 0.85cm, clip=true, width=0.47\textwidth]{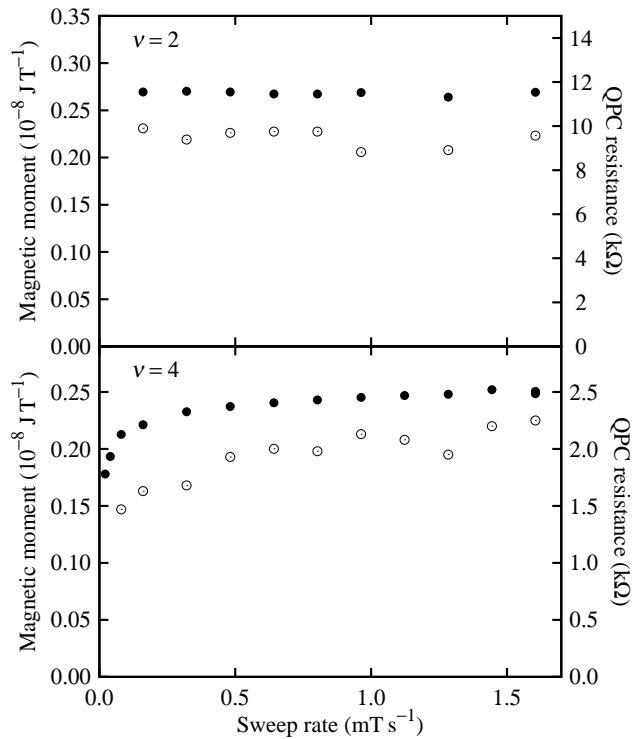}

\caption{Sweep-rate dependences of the induced currents at $\nu=2$ (top) and $\nu=4$ (bottom) at $100\,\mathrm{mK}$. The magnetic moment and QPC resistance were measured simultaneously. Filled symbols are magnetic moment, open symbols are QPC resistance.  The induced currents saturate at high sweep rates. At the slowest sweep rate accessible, the $\nu=2$ induced current remained in the saturated regime.\label{fig:sweep_rate_IV}}

\end{figure}
\subsection{Sweep-rate dependence}\label{sub:sweeprate}
Figure~\ref{fig:sweep_rate_IV} shows the sweep-rate dependences
of the magnetic moment and the hysteretic QPC resistance features at $\nu=2$ and 4.
These measurements are effectively current--voltage ($I$--$V$) curves
for the induced currents: the sweep rate is proportional to the electromotive force
around the perimeter of the 2DES, and the magnetic moment or QPC resistance
is proportional to the induced current. Both $I$--$V$ curves show a saturation of
the induced current at magnetic-field sweep rates greater than $\sim 0.5\,\mathrm{mT\, s^{-1}}$. For $\nu=2$, saturation occurs at the lowest sweep rate used, $0.32\,\mathrm{mT\, s^{-1}}$,
which corresponds to an electromotive force of $1.35\,\mathrm{nV}$. For $\nu=4$
a more gradual reduction in the induced current is observed. In both cases the
shapes of the $I$--$V$ curves derived from the two simultaneous measurements
are the same within experimental error. 
\begin{figure}
\includegraphics[clip=true, width=0.47\textwidth]{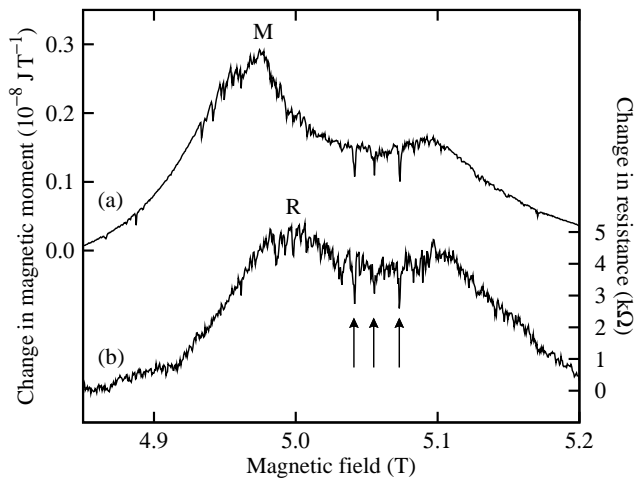}

\caption{The induced current at $\nu=2$, $100\,\mathrm{mK}$, measured using a slow sweep rate ($80\,\mathrm{\mu T\, s^{-1}}$).  Arrows indicate three individual breakdown events which are
correlated between the two types of simultaneous measurement, magnetic moment (upper trace) and QPC resistance (lower trace) -- a clear demonstration that the two measurements share a common physical origin.  The peaks labelled M and R are discussed in the text.\label{fig:noise}}

\end{figure}
\subsection{Noise}\label{sub:noise}
Figure~\ref{fig:noise} shows the induced current feature for $\nu=2$ for both
magnetic moment and QPC resistance using a very slow sweep rate of $80\,\mathrm{\mu T\, s^{-1}}$. Under these
conditions previous investigators \cite{Elliott2006} observed a qualitatively reproducible `noise' structure
around the induced current peaks which they attributed to local QHE breakdown
events occurring at various positions around the perimeter of the 2DES.
In Fig.~\ref{fig:noise}  several `noise' features (e.g. those marked by arrows)
appear in both measurements providing the strongest evidence to date that
the two effects have a common cause. The fact that the noise features have similar sizes in both measurements lends support to the suggestion\cite{Klaffs2004} that the induced current forms one loop around the entire 2DES, rather than many loops localised by impurities. 

Although the noise features occur
at the same magnetic fields, the peaks of the induced current features (peak M in the magnetic moment and peak R in the QPC resistance) do not coincide, as noted earlier in Section~\ref{sub:temp}.
The QPC resistance peak R is shifted to higher magnetic field than the magnetic moment peak M. This clearly demonstrates that the QPC measurement is not simply measuring the size of the induced current (as the magnetic moment measurement does) but is also sensitive to other factors. Evidently
the QPC conductance is influenced by the proximity of the induced current to the QPC as well
as its size, as suggested in Section \ref{sub:simultaneous} above. This proximity is related to the magnetic length. Therefore features in the QPC resistance will tend to shift to higher magnetic field compared with those in the magnetic moment -- a higher field results in a smaller magnetic length, and hence an induced current closer to the QPC.
\begin{figure}
\includegraphics[clip=true, width=0.47\textwidth]{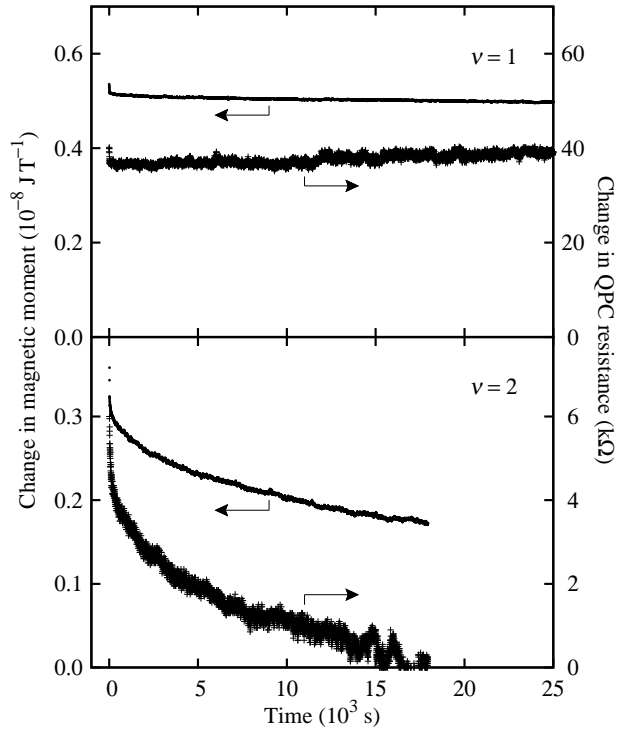}

\caption{Simultaneous measurement of the decays of the induced current, ({\tiny$^\bullet$}) from magnetic moment, and ({\tiny$^+$}) from QPC resistance, at $100\mathrm{\, mK}$. Top: at $\nu=1$; bottom: at $\nu=2$.  The zeroes of the decays are determined by sweeping through the induced current features in both directions, before and after the decay. A significant amount of drift, in the opposite direction to the decay, is evident in the QPC resistance at $\nu=1$.\label{fig:decay}}

\end{figure}
\subsection{Decays}\label{sub:decays}
Figure~\ref{fig:decay} shows the decay of the induced current when the magnetic field is swept to filling factors $\nu=1$ and $2$ and then held constant, measured simultaneously from its magnetic moment and its effect on QPC resistance. At $\nu=1$ both the magnetic moment and the QPC resistance show a rapid initial decay lasting approximately $20\,$s, followed by a much longer period of  slower-than-exponential decay. The QPC resistance also appears to exhibit some drift in the opposite direction to the decay. At $\nu=2$ the magnetic moment decay behaves in the same way as at $\nu=1$ but for the QPC resistance the slow-decay regime is replaced by a faster approximately linear decay.  At $\nu=4$ (not shown) the decay of both the magnetic moment and the QPC resistance is rapid and appears to be exponential. The trend towards faster decays at higher filling factors is due to the decrease in Landau-level separation, $\hbar\omega_{c}$ (where $\omega_{c}=eB/m^*$), as $\nu$ increases.  We note here that although $\nu=1$ is a spin-split feature (and hence the energy between the highest occupied and the lowest unoccupied Landau level is $g\mu_{B}B$), the process by which the decay occurs probably involves tunnelling, which one would expect to conserve spin -- thus the levels involved in the tunnelling are separated by an energy $\hbar\omega_{c}$, not $g\mu_{B}B$. We attribute the difference in behaviour of the long-time decays at $\nu=2$ to the sensitivity of the QPC resistance to the proximity of the induced current as well as its size -- as the induced current decays, the Hall electric field supporting it is also reduced and the induced current becomes more spread out into the bulk of the 2DES.  This effect is not apparent at $\nu=1$ because the current does not decay sufficiently for spreading to become noticeable.
\section{Discussion and Conclusions}
The observations discussed above show that induced currents cause the hysteretic features in both the magnetic moment and the QPC resistance.  We have also investigated the reverse effect, whether the QPC can be used to influence the induced current.  For instance, during an induced-current decay, the bias on the QPC split gate was repeatedly changed to alternately pinch off the QPC and then open it to the point where the 2DES number density under the gate matched that in the bulk of the device. The expectation was that this would cause some extra dissipation and hence increase the decay rate.  No such effect was observed.  In another experiment different voltages were applied  to the two sides of the split gate, with the aim of setting up counter-propagating induced currents in close proximity, again with the expectation of enhanced dissipation. None was apparent. Further studies are necessary to determine how induced currents can be controlled by electrostatically defined nanostructures.

To conclude, in this paper we have demonstrated  that the hysteretic features observed in the magnetoresistance of a QPC and the hysteretic features detected in the magnetic moment of the 2DES surrounding the device have a common origin -- they are caused by long-lived induced currents which occur in the dissipationless regime of the quantum Hall effect.  We have shown this by making simultaneous measurements of the two effects on the same QPC device, and correlating the features, their temperature and sweep-rate dependences and their decays.  We have also identified common features in the reproducible `noise' structure observed at low sweep rates.  It is clear that induced currents are responsible for both effects; however a close comparison of the two measurements suggests that the hysteretic magnetoresistance of the QPC is sensitive not only to the size of the induced currents but also to their proximity to the QPC.

\acknowledgments{The authors would like to thank K. White for constructing the torsion-balance magnetometer, and M. Elliott and R.J. Nicholas for helpful discussions. This work was supported by EPSRC.}

\bibliography{bibliography}

\end{document}